\documentclass[12pt,a4paper]{article}
\usepackage{epsfig}
\usepackage{amssymb}

\newcommand{\beq}{\begin{equation}}
\newcommand{\eeq}{\end{equation}}
\newcommand{\bea}{\begin{eqnarray}}
\newcommand{\eea}{\end{eqnarray}}

\newcommand{\wh}[1]{\widehat{#1}}
\newcommand{\wt}[1]{\widetilde{#1}}
\newcommand{\Tr}{{\rm Tr}}
\newcommand{\cW}{{\cal W}}
\newcommand{\oD}{\overline{D}}
\newcommand{\onab}{\overline{\nabla}}
\newcommand{\dalp}{\dot{\alpha}}
\newcommand{\dbe}{\dot{\beta}}

\addtocounter{page}{-1}

\begin{document}

\begin{flushright}
hep-th/0306035
\end{flushright}

\vskip 2cm

\begin{center}
\Large \bf Chiral rings, anomalies and loop equations\\
in ${\cal N}=1^*$ gauge theories
\end{center}

\vskip 1cm

\begin{center}
Taichi Itoh
\end{center}

\begin{center}
\it BK21 Physics Research Division and Institute of Basic Science\\ 
Sungkyunkwan University, Suwon 440-746, Korea
\end{center}

\vskip 1cm

\begin{abstract}
We examine the equivalence between the Konishi anomaly equations and the 
matrix model loop equations in ${\cal N}=1^*$ gauge theories, the mass 
deformation of ${\cal N}=4$ supersymmetric Yang-Mills. We perform the 
superfunctional integral of two adjoint chiral superfields to obtain an 
effective ${\cal N}=1$ theory of the third adjoint chiral superfield. 
By choosing an appropriate holomorphic variation, the Konishi anomaly 
equations correctly reproduce the loop equations in the corresponding 
three-matrix model. We write down the field theory loop equations explicitly 
by using a noncommutative product of resolvents peculiar to ${\cal N}=1^*$ 
theories. The field theory resolvents are identified with those in the 
matrix model in the same manner as for the generic ${\cal N}=1$ gauge 
theories. We cover all the classical gauge groups.  In $SO/Sp$ cases, 
both the one-loop holomorphic potential and the Konishi anomaly term 
involve twisting of index loops to change a one-loop oriented diagram to 
an unoriented diagram. The field theory loop equations for these cases 
show certain inhomogeneous terms suggesting the matrix model loop equations 
for the ${\bf RP}^2$ resolvent.
\end{abstract}

\vspace*{\fill}
\hrule

\vskip 0.2cm
\noindent
Email: taichi@newton.skku.ac.kr

\baselineskip=18pt
\newpage

\renewcommand{\theequation}{\arabic{section}\mbox{.}\arabic{equation}}
\setcounter{equation}{0}
\section{Introduction}

It was conjectured by Dijkgraaf and Vafa that the glueball superpotentials  
of a large class of ${\cal N}=1$ supersymmetric gauge theories are obtained 
by computing planar diagrams in certain bosonic matrix models 
\cite{DV1,DV2,DV3}. 
The field theory proof of this conjecture was given in \cite{DV3,DGLVZ} 
to show that the relevant contribution to the F-term comes from the planar 
diagrams with two gaugino insertions for each loop. In \cite{DGLVZ}, they 
performed field theory perturbation by starting from the effective 
holomorphic action obtained by path integrating out all the 
anti-chiral superfields. 

The reason why only planar diagrams are relevant to the F-term was clarified 
in \cite{CDSW} based on some properties of chiral operators. 
The chiral ring structure ensures that all the gauge invariant operators are 
equivalent to the operators with at most two gauge field insertions. 
Furthermore, the large $N$ factorization in the matrix model is simply the 
result of  cluster decomposition of correlation functions of chiral operators. 
Then it was shown in \cite{CDSW} that the loop equations in a bosonic matrix 
model are reproduced as the Konishi anomaly equations induced by a certain 
holomorphic variation in the corresponding ${\cal N}=1$ gauge theory. 
In \cite{CDSW}, the equivalence between the matrix model loop equations 
and the generalized Konishi anomaly equations was studied in an ${\cal N}=1$ 
theory with a polynomial superpotential for one adjoint 
matter field.\footnote{
In the very beginning of a literature on Dijkgraaf-Vafa conjecture, 
the idea of using the Konishi anomaly equations to derive the exact 
glueball superpotential appeared in \cite{gorsky} even including 
${\cal N}=1^*$ gauge theories. The relation to the matrix model loop 
equations was recognized in \cite{CDSW}. }
The inclusion of fundamental matter fields was done in \cite{seiberg}. 

Meanwhile, in matrix model side there appeared some attempts to examine the 
conjecture in other kinds of ${\cal N}=1$ gauge theories. In \cite{DV3}, 
the matrix model computation of glueball superpotential was performed in the 
three-matrix model corresponding to the ${\cal N}=1^*$ gauge theories, 
the mass deformation of ${\cal N}=4$ super Yang-Mills. The peculiarity of 
${\cal N}=1^*$ theories is that the theory contains three adjoint chiral 
superfields which interact through a commutator coupling. 
Since the tree-level superpotential is not simply a polynomial, the special 
geometry is implicit in the matrix model loop equations. 
As far as one intends to compute the glueball superpotential in perturbation 
expansion, one can treat the three adjoint fields equally and there is nothing 
different from a generic ${\cal N}=1$ theory with a polynomial superpotential
\cite{DGKV}. However, to obtain the matrix model loop equations, 
one has to first integrate out two of three adjoint matrices to reduce the 
original three-matrix model into an effective one-matrix model. Then loop 
equations are derived as Schwinger-Dyson equations for the matrix model 
resolvents of a surviving matrix \cite{DV3}. 
Motivated by the aforementioned works, it is natural to ask how the Konishi 
anomaly equations in ${\cal N}=1^*$ gauge theories reproduce the matrix 
model loop equations.

The string dual of ${\cal N}=1^*$ gauge theories was studied in \cite{PS}. 
The mass deformation in field theory corresponds to turning on a supergravity 
background in $AdS_5 \times S^5$ compactification. Consequently, D3-branes 
polarize into a noncommutative sphere to form a dielectric D5-brane 
\cite{PS}. This situation is quite similar to the string theory setup of 
Dijkgraaf-Vafa conjecture. 
In a generic ${\cal N}=1$ theory equipped by a polynomial superpotential, 
D5-branes are wrapping around the blownup $S^2$'s obtained by 
resolving the conical singularities of a Calabi-Yau 3-fold described by 
the polynomial superpotential. In ${\cal N}=1^*$, the blownup $S^2$'s get 
replaced with fuzzy spheres resolving the $AdS_5$ singularity to yield the 
dielectric D5-branes.

In this paper, we derive the Konishi anomaly equations in ${\cal N}=1^*$ 
gauge theories. Instead of integrating out anti-chiral superfields 
\cite{DGLVZ}, we first path integrate out two adjoint chiral superfields and 
their anti-chiral partners to gain an effective single adjoint theory. We also 
have corrections to the Kahler potential but these are irrelevant to the 
Konishi anomaly equations. Due to the ${\cal N}=1$ non-renormalization 
theorem, there is no perturbative corrections to the tree-level superpotential
so that the lowest correction should include at least two spinor gauge 
superfields. Then using the chiral ring property, we will show the equivalence 
between the Konishi anomaly equations and the matrix model loop equations 
under an appropriate identification of matrix model resolvents in the field 
theory. We first derive the field theory loop equations in $U(N)$ gauge 
theories and check the consistency of results with \cite{DV3}. The field 
theory loop equations are explicitly written down by using a noncommutative 
product of resolvents peculiar to ${\cal N}=1^*$ theories. Then we 
extend the results to cover $SO/Sp$ cases also. Especially, we extract the 
loop equation for the matrix model resolvent on ${\bf RP}^2$ out of the 
Konishi anomaly equations.

There are some remarks on the connection between the field theory 
perturbation and the matrix model perturbation. 
As noted in \cite{DHKS1,KS}, when we evaluate the glueball superpotential 
via matrix model perturbation, there arises a discrepancy between the matrix 
model and the field theory results beyond $h$-loops, with $h$ denoting the 
dual Coxeter number of the gauge group. This is due to the mixing ambiguities 
of chiral operators which have been discussed earlier in \cite{DKM,ADK}. 
In \cite{AIVW} they proposed one way to avoid these ambiguities through 
embedding the gauge group into a certain supergroup. 
However, the loop equations in matrix models are obtained by summing up 
all planar diagrams and we do not have to worry about the subtlety of 
order by order computation of planar diagrams. 
Instead, we have to identify the matrix model resolvents with those in 
the field theory. This was done in \cite{KRS} in the generic ${\cal N}=1$ 
theories. The validity of this identification in ${\cal N}=1^*$ was 
suggested in \cite{PTZ}. In this paper, we will check the identification by 
deriving the field theory loop equations explicitly.

The organization of this paper is as follows. In section 2, we sketch 
the F-term computation of the one-loop effective theory, leaving 
the calculation details in appendix A. In section 3 we discuss Konishi 
anomaly equations in ${\cal N}=1^*$ and identify the field theory 
resolvents in the corresponding matrix model. The extension to $SO/Sp$ 
gauge groups is also discussed. Section 4 is devoted to conclusions and 
discussions. 

\begin{figure}
\begin{center}
\centerline{\hbox{\psfig{file=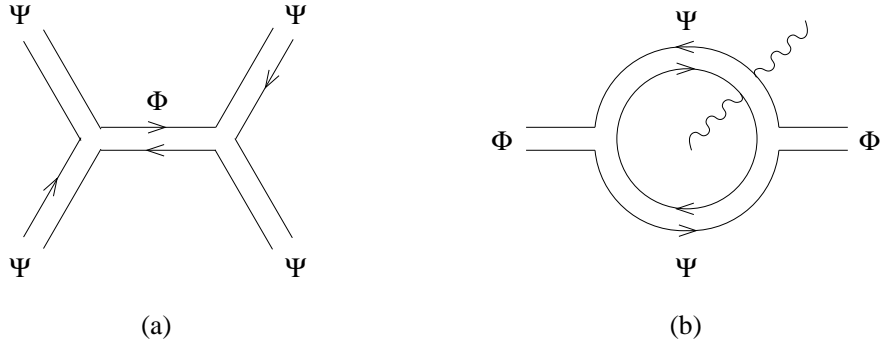,height=4.5cm}}}
\caption{Two ways to reduce a two adjoint theory to a single adjoint theory;
(a) Integrating $\Phi$ field to obtain ${\rm Tr}\Psi^4$, (b) integrating 
$\Psi$ field to obtain ${\rm Tr}(W^\alpha \Phi^2){\rm Tr}W_\alpha$.}
\label{fig:twoadj}
\end{center}
\end{figure}

\setcounter{equation}{0}
\section{One-loop effective action in ${\cal N}=1^*$ gauge theories}

\subsection{The meaning of one-loop effective action}

Let us start with the discussion of an ${\cal N}=1$ gauge theory with 
two adjoint chiral superfields. The tree-level superpotential is given by
$$
W(\Phi, \Psi)={\rm Tr}(m \Phi^2 +\Phi \Psi^2) 
$$
as in footnote 1 in \cite{CDSW}. There are two ways to reduce this 
theory to a single adjoint theory. One is to integrate out $\Phi$ first. 
The only relevant diagram is the tree-diagram in figure \ref{fig:twoadj} 
(a) and the theory turns out to be a single adjoint theory with the 
tree-level superpotential $W(\Psi)=-{\rm Tr}\Psi^4/4m$. The other is to 
integrate $\Psi$ first, then the effective theory is given by one-loop 
diagrams as in figure \ref{fig:twoadj} (b). For each diagram not to vanish, 
it must contain at least two gauge field insertions as shown in the figure. 
This is consistent with the counting rule given in \cite{DGLVZ,CDSW}, 
each loop has exactly two gluino insertions; two for one of two 
index loops or one for each index loop. Therefore starting from this 
effective theory, the Konishi anomaly equations seem to reproduce the 
matrix model loop equations. However, in this simplest case, we do not 
have to choose the latter way rather than manipulate the simple one-adjoint 
theory given in the former procedure. 

The ${\cal N}=1^*$ theory contains three adjoint chiral superfields 
with three-point vertex involving a commutator of two adjoints. To gain 
the single adjoint theory, one must integrate out two of them and the 
situation is similar to the case (b). In the rest of this section, 
we will sketch the computation of the effective action in ${\cal N}=1^*$ 
gauge theories. For the details of computation, see appendix A.

\subsection{Reduction to the ${\cal N}=1$ effective single adjoint theory}

Using covariantly constrained chiral superfields, the lagrangian for 
an ${\cal N}=1^*$ gauge theory is written as\footnote{
Throughout this paper, we will use the notation of \cite{lykken} 
with Minkowski metric $(+---)$.}
\bea
{\cal L} &=& {1 \over g^2} \Biggl[
\int d^4 \theta \;\Tr\left(\overline{\Phi}_1\Phi_1
+\overline{\Phi}_2\Phi_2+\overline{\Phi}_3\Phi_3\right) 
+\int d^2 \theta\,{1 \over 2}\,W(\Phi)
+\int d^2 \bar{\theta}\,{1 \over 2}\,\overline{W}(\overline{\Phi})\Biggr]
\nonumber\\ &&
+{1 \over 32\pi}{\rm Im}
\left[ \tau_0 \int d^2 \theta\; \Tr\, W^\alpha W_\alpha \right],
\label{onestarL}
\eea
where the tree-level superpotential is given by
\beq
W(\Phi)={\rm Tr}\left[ 2\,\Phi_3 \,[\Phi_1,\Phi_2]
+m\,(\Phi_1^2+\Phi_2^2+\Phi_3^2) \right],
\eeq
and $\tau_0$ is the bare Yang-Mills coupling
$$
\tau_0 = {\theta \over 2\pi}+{4\pi i \over g^2}.
$$
We will sketch our one-loop computation in $U(N)$ gauge groups. 
In $SO/Sp$ cases, the chiral superfields $\Phi_i$ can be either symmetric 
or antisymmetric tensor fields so that one has to insert appropriate 
projection operators to do trace calculation correctly. However this can 
be easily done through a small modification of $U(N)$ results. 

In the lagrangian (\ref{onestarL}), $\Phi_i$ and $\overline{\Phi}_i$ are 
gauge covariant superfields subject to the constraints
$$
\overline{\nabla}_{\dot{\alpha}}\Phi_i =0,\quad 
\nabla_\alpha \overline{\Phi}_i=0.
$$
In the gauge chiral representation, they are related to the chiral
superfields $\Psi_i$, $\overline{\Psi}_i$ via
$$
\Phi_i = \Psi_i, \quad
\overline{\Phi}_i = e^{-{\rm ad}(V)}\overline{\Psi}_i 
\equiv e^{-V}\,\overline{\Psi}_i\, e^V,
$$
and accordingly the gauge covariant spinor derivatives are defined 
by \cite{GGRS}
$$
\overline{\nabla}_{\dot{\alpha}} = \overline{D}_{\dot{\alpha}},\quad
\nabla_\alpha = e^{-{\rm ad}(V)}D_\alpha e^{{\rm ad}(V)}
\equiv {\rm ad}\!\left(e^{-V}D_\alpha e^V\right).
$$
Note that in this representation $\Phi_i$ are ordinary chiral superfields 
eliminated by $\overline{D}_{\dot{\alpha}}$. The gauge field strength 
superfield $W_\alpha$ is defined by
$$
W_\alpha =-{1\over 4}(\overline{D}^2 e^{-V} D_\alpha e^V),
$$
where the bracket means that the spinor derivatives are active only 
inside it.

Though the interaction involves a commutator of two adjoint superfields, 
it is still quadratic in $\Phi_2$ and $\Phi_3$ so that one can path 
integrate out $\Phi_2$, $\Phi_3$ and their anti-chiral partners to obtain 
the perturbatively exact effective action for an ${\cal N}=1$ gauge theory 
with a single adjoint field $\Phi\equiv\Phi_1$. Each diagram depends on 
the external gauge field in gauge covariant way through the covariant 
superfield propagators. It is further expanded into the diagrams with 
external gauge fields, namely $W_\alpha$'s, by using the Schwinger-Dyson 
equations for superfield propagators. One can easily compute the 
super-Feynman integrals by using the so-called covariant `D-algebra' 
\cite{GGRS,GZ}. Consequently, each of the one-loop diagrams reduces to 
a gauge-invariant non-local vertex among external $\Phi$ and $W_\alpha$ 
legs. 

The most important point in the field theoretical proof of Dijkgraaf-Vafa 
conjecture might be the chiral ring property of chiral operators, i.e, 
the equivalence of gauge invariant operators modulo chiral rings. 
It ensures that the planarity of computation of the glueball superpotentials. 
As explained in \cite{CDSW}, the correlation functions of gauge-invariant 
chiral operators are independent of the spatial coordinates and thereby 
factorized into a product of one-point functions due to the cluster 
decomposition of correlation functions. One can therefore reduce the 
nonlocal vertices of one-loop diagrams to a local holomorphic potential 
by extracting only the lowest order terms in the derivative expansion.

According to the ${\cal N}=1$ non-renormalization theorem, the terms 
corresponding to the one-loop diagrams without gauge field legs become 
trivially zero so that the non-trivial terms are those with at least 
two gauge field legs. Since, within the equivalence modulo chiral rings, 
$\Phi$ commutes with $W_\alpha$ and $W_\alpha$ obeys Fermi statistics, 
namely \cite{CDSW}
$$
[ \Phi, W_\alpha ]=0, \quad \{ W_\alpha, W_\beta \}=0,
$$
one may only compute the diagrams with two gauge field legs as well as even 
number of $\Phi$ legs\footnote{
It is topologically impossible to create a one-loop diagram with 
odd number of $\Phi$ legs by contracting the cubic interaction vertices 
in ${\cal N}=1^*$ gauge theories; An external $\Phi(=\Phi_1)$ leg couples 
to an internal $\Phi_2$ line as well as an internal $\Phi_3$ line.} 
to obtain the F-term of effective single adjoint theory.
Therefore the lagrangian what we expect for the effective ${\cal N}=1$ 
theory can be written as
\beq
{\cal L}_{\rm eff} = \int\! d^4 \theta\,K(\overline{\Phi},\Phi) 
+\left[\int \!d^2\theta \,{1 \over 2}\,\wt{W}(\Phi,W_\alpha) + h.c. \right]
+{1 \over 32\pi}{\rm Im}
\left[\tau_0 \int d^2 \theta\;\Tr\left(W^\alpha W_\alpha\right) \right].
\label{effL}
\eeq
Here we think of the theory as already renormalized, having all the 
one-loop corrections to the gauge coupling. The UV behavior of the theory 
is the same as ${\cal N}=4$ super Yang-Mills so that there is no one-loop 
correction to the gauge kinetic term at all. Although the Kahler potential 
contains one-loop corrections, they are irrelevant to the Konishi anomaly 
equations supposed the vacuum is supersymmetric. 

\subsection{Computation of one-loop holomorphic interactions}

We are now led to compute the holomorphic $2n$-point function with $2n$ 
$\Phi$ fields as well as two spinor gauge fields. After some straightforward 
calculations, the full holomorphic corrections modulo chiral rings are 
summarized to be a holomorphic potential
\bea
\widetilde{W}(\Phi,W_\alpha) &=& 
{m \over g^2}{\rm Tr}\,\Phi^2 
+\sum_{n=1}^{\infty}\,{(-)^n \over n(2n+1)m^{2n}}\,
W^{(2n)}(\Phi,W_\alpha),\nonumber\\
W^{(2n)}(\Phi,W_\alpha) &=& W^{(0,2n)}(\Phi,W_\alpha) 
+{1 \over 2n}\sum_{r=1}^{2n-1}\sum_{s=1}^{2n-r} W^{(s,2n-s)}
(\Phi,W_\alpha),
\label{holom}
\eea
where $W^{(0,2n)}$ and $W^{(s,2n-s)}$ $(s=1,\dots,2n-1)$ are given by 
\bea
W^{(0,2n)}(\Phi,W_\alpha) &=& {1\over 32\pi^2} {\rm Tr}_{\rm ad}\! 
\left[ \cW^\alpha \cW_\alpha \phi^{2n} \right], \nonumber\\
W^{(s,2n-s)}(\Phi,W_\alpha) &=& {1\over 32\pi^2} {\rm Tr}_{\rm ad}\! 
\left[ \cW^\alpha \phi^s \cW_\alpha \phi^{2n-s} \right].
\label{holom2}
\eea
Here the traces are taken over the gauge adjoint representation. Note that 
there arises $\phi\equiv{\rm ad}(\Phi)$, instead of $\Phi$, inside the traces. 
They are rewritten as gauge fundamental traces of nested commutators;
\bea
{\rm Tr}_{\rm ad}\!\left[ \cW^\alpha \cW_\alpha \phi^{2n} \right] 
&=& {\rm Tr}\,\Bigl[ e_{ji} \{ W^\alpha, [ W_\alpha, 
[ \underbrace{ \Phi, ..... [ \Phi }_{2n}, e_{ij} ] ... ] ] \} \Bigr],
\nonumber\\
{\rm Tr}_{\rm ad}\!\left[ \cW^\alpha \phi^s \cW_\alpha \phi^{2n-s} \right]
&=& {\rm Tr}\,\Bigl[ e_{ji} \{ W^\alpha, 
[ \underbrace{ \Phi, ..... [ \Phi}_{s}, [ W_\alpha, 
[ \underbrace{ \Phi, ..... [ \Phi}_{2n-s}, e_{ij} ] ... ]]] ...] \} \Bigr],
\label{adjTr}
\eea
where $(e_{ij})_{kl}=\delta_{ik}\delta_{jl}$ form a basis of gauge adjoint 
matrices to satisfy $X = e_{ij} X_{ij}$ for any gauge adjoint 
matrix $X$ whose $(i,j)$-component is $X_{ij}$. One can expand the nested 
commutators into operator products by using the identity;
$$
[ \underbrace{\Phi, ..... [ \Phi}_n, X ] ... ]
=\sum_{r=0}^n \left({n \atop r}\right)(-)^r \,\Phi^{n-r}X\,\Phi^r.
$$

\begin{figure}
\begin{center}
\centerline{\hbox{\psfig{file=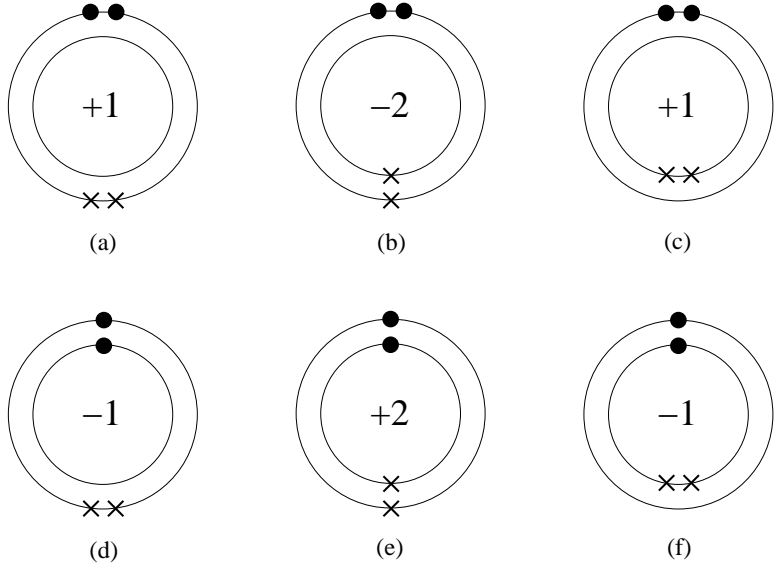,height=7.5cm}}}
\caption{The holomorphic 2-point functions with two insertions of 
external gauge fields. A cross (dot) denotes an external $\Phi$ ($W_\alpha$) 
leg inserted to either of two index loops. For instance, the diagram (d) 
corresponds to $-{\rm Tr}(\Phi^2 W^\alpha){\rm Tr}\,W_\alpha$.}
\label{fig:oneloop}
\end{center}
\end{figure}

Eventually, the holomorphic functions are given by the summations of double 
traces each of which corresponds to a one-loop ribbon diagram with two 
index loops.
\bea
W^{(0,2n)}(\Phi,W_\alpha) \!&=&\! 
{1 \over 16\pi^2}\sum_{r=0}^{2n}\left({2n \atop r}\right)(-)^r 
\nonumber\\&& 
\Bigl[\Tr\,\Phi^r \,\Tr(\Phi^{2n-r}W^2)
-\Tr(\Phi^r W^\alpha)\,\Tr(\Phi^{2n-r}W_\alpha)\Bigr],\nonumber\\
W^{(s,2n-s)}(\Phi,W_\alpha) \!&=&\!
{1 \over 16\pi^2}\sum_{r=0}^s \sum_{p=0}^{2n-s}
\left({s \atop r}\right)\left({2n-s \atop p}\right)(-)^{r+p}
\nonumber\\&& 
\Bigl[\Tr\,\Phi^{r+p} \,\Tr(\Phi^{2n-r-p}W^2)
-\Tr(\Phi^{r+p} W^\alpha)\,\Tr(\Phi^{2n-r-p}W_\alpha)\Bigr].
\label{doubleTr}
\eea
The binomial factors in the summations count the different ways to 
distribute $2n$ external $\Phi$ legs either on the inner index loop or 
on the outer index loop. The $\Phi$ legs on the same index loop commute 
with each other, while two $\Phi$ legs on different index loops are 
noncommutative. This is due to the commutator coupling in the original 
${\cal N}=4$ super Yang-Mills. In figure \ref{fig:oneloop}, we showed 
the ribbon diagrams corresponding to the double traces in $W^{(0,2)}$. 
In the diagram (d), the inner index loop has one insertion of 
$W^\alpha$, while the outer index loop contains two $\Phi$ legs as well as
$W_\alpha$ to yield the double trace of 
$-{\rm Tr}\,W^\alpha\,{\rm Tr}(\Phi^2 W_\alpha)$.

\setcounter{equation}{0}
\section{Konishi anomaly equations in ${\cal N}=1^*$ theories}

\subsection{Field theory resolvents and holomorphic interactions}

We will see shortly that what one needs for evaluating the Konishi anomaly 
equations is the $\Phi$ derivative of the holomorphic function 
(\ref{holom}). 
In spite of the multi-summations in (\ref{holom}), they are summable by 
using the field theory resolvents, i.e, the generating functions for the 
gauge invariant chiral operators. Within the equivalence modulo chiral 
rings, all of those operators reduce to the three classes 
of operators;
$$
\Tr\,\Phi^r, \quad \Tr(\Phi^r W_\alpha), \quad \Tr(\Phi^r W^2). 
$$
Thus we are led to introduce the field theory resolvents;
\bea
R(z) &\equiv& \Tr\,\wh{R}(z) \;=\; 
-{1 \over 32\pi^2}\Tr\, W^2 \,{1\over z-\Phi},\nonumber\\
w_\alpha(z) &\equiv& \Tr\,\wh{w}_\alpha(z) \;=\;
{1 \over 4\pi}\Tr\, W_\alpha \,{1\over z-\Phi},\nonumber\\
T(z) &\equiv& \Tr\,\wh{T}(z) \;=\; \Tr\,{1\over z-\Phi}.
\eea

Suppose that $\Phi$ has some continuous distributions of its eigenvalues 
around the classical vacua. Then by choosing the counter $\cal C$ 
on the complex $z$-plane such that it encircles the branch cuts of 
eigenvalue distributions, the gauge invariant chiral operators are 
given by the contour integrals of field theory resolvents;
\bea
{\rm Tr}(\Phi^r W^2) &=& \oint_{\cal C}{dz \over 2\pi i}\,z^r R(z),
\nonumber\\
{\rm Tr}(\Phi^r W_\alpha) &=& \oint_{\cal C}{dz \over 2\pi i}\,
z^r w_\alpha(z), \nonumber\\
{\rm Tr}\,\Phi^r &=& \oint_{\cal C}{dz \over 2\pi i}\,z^r T(z).
\label{cont}
\eea
These formulas convert the power of $\Phi$ into the power of $z$ 
free from the traces of resolvents and make the summations involving 
binomial coefficients doable. 

Consequently, we arrive at a rather simple result;
$$
{\partial \over \partial\Phi} W^{(s,2n-s)}(\Phi,W_\alpha)
= 4n \oint_{\cal C}{dz \over 2\pi i}\,(z-\Phi)^{2n-1}
\left[R(z)+w^\alpha(z)\wh{v}_\alpha +T(z)\wh{S} \right],
$$
for $s=0,1,\dots,2n-1$. Here we have introduced the chiral operators
$$
\wh{S} \equiv -{1 \over 32\pi^2}W^2, \quad
\wh{v}_\alpha \equiv {1 \over 4\pi}W_\alpha.
$$
Note that we attach a hat symbol to a gauge covariant chiral operator 
whose trace yields the corresponding gauge singlet chiral field or 
the field theory resolvent.\footnote{
For instance, $\wh{S}$ denotes a chiral operator whose trace 
$S\equiv\Tr\,\wh{S}$ becomes a glueball superfield for the traceful 
gauge group $U(N)$ and is different from $\wh{S}$ in \cite{CDSW}.} 
Together with the factor $4n$ above, the summations 
in the second equation of (\ref{holom}) provide a factor $2n(2n+1)$ 
and cancels the prefactor $1/n(2n+1)$ in the first equation of 
(\ref{holom}). 
Finally the $\Phi$ derivative of $\wt{W}(\Phi,W_\alpha)$ is determined as
\bea
\widetilde{W}^{\,\prime}(\Phi,W_\alpha) &=&
{2m \over g^2}\Phi -\oint_{\cal C} {dz \over 2\pi i} 
\left[ {1 \over z - \Phi + im} + {1 \over z - \Phi - im} \right]
\nonumber\\&&\;
\left[R(z)+w^\alpha(z)\wh{v}_\alpha +T(z)\wh{S} \right].
\label{derW}
\eea

\subsection{Konishi anomaly as the field theory loop equations}

The Konishi anomaly equations are obtained as anomalous Ward-Takahashi 
identities associated with an arbitrary holomorphic variation 
$\delta\Phi = f(\Phi,W_\alpha)$, while keeping $\overline{\Phi}$ unchanged, 
in the effective single adjoint theory (\ref{effL}). Specifically, 
they are given as an identity
$$
0 = \int \!d^8 z^{\prime} \!\int \,[d\overline{\Phi}][d\Phi] \;\Tr_{\rm ad} 
\left[ {\partial \over \partial \Phi(z^{\prime})}f(\Phi(z),W_\alpha(z))
\exp\left(i\!\int \!d^4 x \,{\cal L}_{\rm eff}\right)\right],
$$
where the trace must be an adjoint trace since inside the square brackets 
is a matrix in the gauge adjoint representation. 
Thus we obtain the generalized Konishi anomaly
\beq
-{1\over 4}\oD^2 \langle J_f \rangle = 
i\left \langle \int \!d^8 z^{\prime} \,\Tr_{\rm ad} \left[ 
{\partial f(\Phi(z),W_\alpha(z)) \over \partial \Phi(z^{\prime})} 
\right] \right \rangle
-{1\over 2}\left \langle \Tr \left[ f(\Phi,W_\alpha)
\wt{W}^{\,\prime}(\Phi,W_\alpha) \right] \right \rangle.
\label{gKonishi}
\eeq
The current $J_f$ is induced by the holomorphic variation of the Kahler 
potential and is given by
\beq
J_f = \Tr\left[\frac{\partial K(\overline{\Phi},\Phi)}
{\partial \Phi}f(\Phi, W_\alpha)\right].
\label{Kahler}
\eeq

Now the Kahler potential might include the one-loop corrections with 
external $\Phi$ and $\overline{\Phi}$ legs.
However, on the supersymmetric vacuum, 
$\overline{D}^2 \langle J_f\rangle =0$ and the only D-term contributions 
through $J_f$ to the Konishi anomaly simply vanishes \cite{CDSW}. 
Therefore we do not have to compute the D-term corrections.
The anomaly term in (\ref{gKonishi}) can be calculated by inserting the 
regulator function $\exp(-L)$ with 
$L=\oD^2 e^{-{\rm ad}(V)}D^2 e^{{\rm ad}(V)}$. 
In our notation, the anomaly coefficient is $1/64\pi^2$ \cite{KoSh} 
and (\ref{gKonishi}) goes to
\beq
-{1\over 32\pi^2} \left \langle \Tr \left[ e_{ij} \left\{ W^\alpha,
\left[ W_\alpha, \frac{\partial f(\Phi,W_\alpha)}{\partial\Phi_{ij}}
\right] \right\} \right]
\right \rangle = \left \langle \Tr \left[ f(\Phi,W_\alpha)
\wt{W}^{\,\prime}(\Phi,W_\alpha) \right] \right \rangle,
\label{gfloop}
\eeq
which is the same as the generalized Konishi anomaly in \cite{CDSW} 
defining the glueball superfield $S$ to have the numerical factor 
$1/32\pi^2$. This is the reason why we inserted the factor $1/2$ 
in front of the superpotential. 

In order to obtain the Konishi anomaly equations as the field theory loop 
equations comparable to those in matrix models, we choose the holomorphic 
variation as the form suggested by the matrix model resolvents, namely
\beq
f(\Phi,W_\alpha) = \wh{\cal R}(z) \equiv \wh{\cal S}{1\over z-\Phi}, 
\label{holovari}
\eeq
where as performed in \cite{CDSW} we have introduced the ${\cal N}=2$ 
glueball operator
$$
\wh{\cal S} \equiv \wh{S} +\psi^\alpha \wh{v}_\alpha 
-{1\over2}\psi^\alpha \psi_\alpha \wh{1}.
$$
The hidden ${\cal N}=2$ supersymmetry is implemented by the 
fermionic coordinates $\psi_\alpha$. Note that the operator $\wh{\cal R}(z)$ 
is decomposed into the ${\cal N}=1$ operators like
$$
\wh{\cal R}(z) = \wh{R}(z) +\psi^\alpha \wh{w}_\alpha(z)
-{1\over 2}\psi^\alpha \psi_\alpha \wh{T}(z),
$$
and therefore the combination of the resolvents in (\ref{derW}) 
can be rewritten as
\beq
R(z) + w^\alpha (z) \wh{v}_\alpha + T(z) \wh{S} 
=-2\int\!d^2\psi\,{\cal R}(z)\,\wh{\cal S}.
\label{fresolv}
\eeq
This ensures that the ${\cal N}=2$ language introduced in \cite{CDSW} 
to describe the field theory prepotential is still available in 
${\cal N}=1^*$ gauge theories.

Substituting the special variation (\ref{holovari}) into 
(\ref{gKonishi}), the generalized Konishi anomaly turns out to be
\beq
{\cal R}(z)^2 = \Tr \left[ \wt{W}^{\,\prime} \wh{\cal R}(z) \right],
\eeq
which can be further decomposed into three equations for the three 
resolvents;
\bea
R(z)^2 &=& \Tr \left[ \wt{W}^{\,\prime} \wh{R}(z) \right],\nonumber\\
2R(z)w_\alpha(z) &=& \Tr \left[ \wt{W}^{\,\prime} 
\wh{w}_\alpha(z) \right],\nonumber\\
2R(z)T(z)+w^\alpha(z)w_\alpha(z) &=& 
\Tr \left[ \wt{W}^{\,\prime}\, \wh{T}(z) \right].
\label{floop}
\eea
Here we have used the factorization property of chiral operator correlation 
functions and suppressed the bracket symbols of field theory vevs. 
The formal expressions of the field theory loop equations (\ref{floop}) 
are the same as those for the ordinary ${\cal N}=1$ theory with a polynomial 
superpotential. The right hand side of (\ref{floop}) seems rather 
complicated  because $\wt{W}^{\,\prime}(\Phi,W_\alpha)$ involves all the 
field theory resolvents through the combination of (\ref{fresolv}).  
However, the loop equations are further simplified by virtue of the chiral 
ring properties;
$$
\wh{S}^2 =0,\quad \wh{S}\,\wh{v}_\alpha =0.
$$

It is straightforward by using this chiral ring properties to write down 
the Konishi anomaly equations (\ref{floop}) explicitly in the field 
theory resolvents. 
\bea
R(z)^2 &=& 
(R\circ R)(z)+[V^{\,\prime}R(z)]_-  \nonumber\\ 
2R(z)w_\alpha(z) &=& 
(R\circ w_\alpha )(z)+(w_\alpha \circ R)(z)
+[V^{\,\prime}w_\alpha(z)]_-, \label{floop2}\\
2R(z)T(z)+w^\alpha(z)w_\alpha(z) &=& 
(R\circ T)(z)+(w^\alpha \circ w_\alpha )(z)+(T\circ R)(z)
+[V^{\,\prime}\,T(z)]_-,\nonumber
\eea
where we have defined a noncommutative product $(O_1 \circ O_2)$ between 
two arbitrary functions $O_1$, $O_2$ by
\beq
(O_1 \circ O_2) (x) \equiv
\oint_{\cal C} {dz \over 2\pi i} \left[ 
{O_1 (z+i)-O_1 (x) \over z-x+i}+c.c. \right] O_2 (z).
\eeq
We also have replaced the mass deformation $\Phi^2$ with 
a polynomial superpotential $V(\Phi)$. $[V^{\,\prime}R(z)]_-$ means 
to drop non-negative powers in the Laurent expansion of 
$V^{\,\prime}(z)R(z)$ \cite{CDSW}.
Here to compare our field theory results with the matrix model 
results in \cite{DV3}, we have rescaled all the superfields 
to be dimensionless;
$$
\Phi \to m\Phi, \quad W_\alpha \to {m^{3/2}\over g}W_\alpha,\quad 
\psi_\alpha \to {m^{3/2}\over g}\psi_\alpha.
$$
This rescaling amounts to make the resolvents to scale correctly 
and yields (\ref{floop2}). 

We notice that the first equation is closed with respect to $R(x)$ and is 
equivalent to the matrix model loop equation through identification of 
$R(x)$ as the resolvent on 2-sphere. To further simply the equations, 
we now introduce the spectral decompositions
\bea
 R(z) &=& \int_{-a}^a d\lambda\,\rho(\lambda)\,
{1 \over z - \lambda},\nonumber\\
 w_\alpha (z) &=& \int_{-a}^a d\lambda\,\rho(\lambda)u_\alpha(\lambda)\,
{1 \over z - \lambda},\nonumber\\
 T(z) &=& \int_{-a}^a d\lambda\,\rho(\lambda)c(\lambda)\,
{1 \over z - \lambda},
\label{spec}
\eea
supposed that the gauge symmetry is unbroken and the eigenvalues of 
the resolvents are distributed along a single branch cut on the complex 
$z$-plane. The spectral densities obey the constraints given by the contour 
integrals of resolvents. By setting $r=0$ in (\ref{cont})  and applying 
the spectral decompositions above, they are written as
$$
\int_{-a}^a d\lambda\,\rho(\lambda) = S, \quad
\int_{-a}^a d\lambda\,\rho(\lambda) u_\alpha (\lambda) = v_\alpha, \quad
\int_{-a}^a d\lambda\,\rho(\lambda) c(\lambda) = N.
$$

By using (\ref{spec}) in (\ref{floop2}),  the Konishi anomaly 
equations provide the following conditions for the resolvents along 
the branch cut $z \in (-a,a)$;
\bea
 2R(z) &=& V^{\prime}(z) + R(z+i) + R(z-i),\nonumber\\
 2w^\alpha (z) &=& w^\alpha (z+i) + w^\alpha (z-i),\nonumber\\
 2T(z) &=& T(z+i) + T(z-i),
\label{floop3}
\eea
These equations fix the analytic properties of resolvents and uniquely 
determine the corresponding elliptic curve of a Riemann surface. 
The first equation reproduces the matrix model result in \cite{DV3,KKN} 
with $V^{\prime}(z)=2z$. It is an equation of motion for a probe matrix 
eigenvalue $z$. The mass deformation of ${\cal N}=1^*$ has been 
generalized to an arbitrary polynomial potential $V(\Phi)$ and the 
corresponding matrix model loop equation was analyzed in 
\cite{DHKS1,DHKS2}. It is exactly the first equation in (\ref{floop3}) with 
$V^{\prime}(z)$ generalizing the mass deformation $2z$ of the original 
${\cal N}=1^*$ theory. 

\begin{figure}
\begin{center}
\centerline{\hbox{\psfig{file=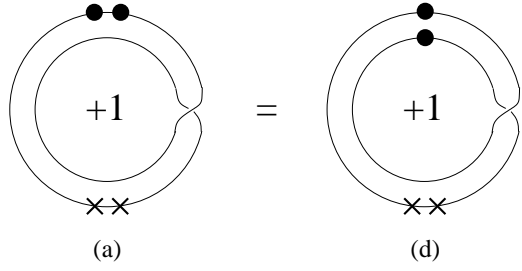,height=3.5cm}}}
\caption{The double trace diagrams in figure \ref{fig:oneloop} transfer to 
single trace diagrams due to the twisted part of $SO/Sp$ projection operator.  
The diagrams (a), (d) in figure \ref{fig:oneloop} turn out to be the same 
single trace diagram of $\Tr(W^2 \Phi^2)$.}
\label{fig:twist}
\end{center}
\end{figure}

\subsection{The field theory loop equations in $SO/Sp$ cases}

The extension to $SO/Sp$ cases is straightforward as worked out 
in \cite{KRS,AC,JO,INO,ACHKR}. We do not have to repeat the computation 
of one-loop effective action performed in $U(N)$ except for the last step to 
calculate the adjoint traces in (\ref{adjTr}). In $SO/Sp$ groups, there 
exist two-index tensor fields $\Phi_{ij}$ either symmetric or antisymmetric 
under $i \leftrightarrow j$. Now the field $\Phi$ does not distinguish two 
index lines of a one-loop ribbon diagram and induces twisted diagrams like 
in figure \ref{fig:twist}. These diagrams cause a bit modification of 
the holomorphic potential $\wt{W}(\Phi,W_\alpha)$.  
Accordingly, the Konishi anomaly equations are changed to involve the 
unoriented geometry of an ${\bf RP}^2$ Riemann surface.  

To see this in more detail, let us introduce the projection operator 
\cite{INO,KRS}
\beq
(P_{ij})_{kl} \equiv P_{ij,kl} = {1 \over 2} \left(\delta_{ik}\delta_{jl}
+\sigma t_{il}t_{jk}\right),
\eeq
where $t_{ij}$ denotes the invariant tensors for $SO/Sp$ groups, namely
\bea
&& t_{ij}=\delta_{ij}, \quad \sigma=-1, \quad 
\mbox{$SO$ antisymmetric ($SO^-$)},\nonumber\\
&& t_{ij}=\delta_{ij},  \quad \sigma=+1, \quad 
\mbox{$SO$ symmetric ($SO^+$)},\nonumber\\
&& t_{ij}=J_{ij}, \quad \sigma=-1, \quad 
\mbox{$Sp$ symmetric ($Sp^-$)},\nonumber\\
&& t_{ij}=J_{ij}, \quad \sigma=+1, \quad 
\mbox{$Sp$ antisymmetric ($Sp^+$)}, 
\eea
where $J_{ij}$ is a symplectic tensor satisfying $J^T =-J$, $J^2 =-\wh{1}$.
The tensor superfield $\Phi$ obeys the projection $\Phi_{ij}=\Tr(P_{ji}\Phi)$ 
so that $P_{ij}$ plays the same role as $e_{ij}$ for $U(N)$ groups. The only 
change in the one-loop calculation is to replace $e_{ij}$ in 
(\ref{adjTr}) with $P_{ij}$. 

The untwisted part of $P_{ij}$ does not cause any change in 
(\ref{doubleTr}) except for the numerical factor 1/2.
The twisted part causes an interchange of two indices and a double trace of 
two index loops turns out to be a single trace of one index loop.
The twisting of indices also transposes the operators on one of two index 
loops. This yields an overall minus sign to the second term in 
(\ref{doubleTr}). Accordingly, for the twisted sector the first and the 
second terms give the same single trace modulo chiral ring equivalence. 

In figure \ref{fig:twist} we demonstrated how twisting occurs in the simplest 
case of $W^{(0,2)}$. For the untwisted sector, we have 
(a) $+\Tr(W^2 \Phi^2)\,\Tr\,\wh{1}$ and 
(d) $-\Tr(W_\alpha\Phi^2)\,\Tr\, W^\alpha$. 
For the twisted sector, by using 
$\Phi^T =\sigma \,t\, \Phi \,t^{-1}$ and 
$W_\alpha^T =-t\, W_\alpha \,t^{-1}$, 
we can see that (a) and (d) transfer to the same single trace;
\bea
+\Tr(W^2 \Phi^2)\,\Tr\,\wh{1} & \to & 
+\sigma\Tr\left[t\,(W^2 \Phi^2)^T \,t \,\wh{1} \right]
\;=\; \pm\sigma\,\Tr(W^2 \Phi^2),\nonumber\\
-\Tr(W^\alpha\Phi^2)\,\Tr\, W_\alpha & \to & 
-\sigma\Tr\left[ t\,(W^\alpha\Phi^2)^T t \,W_\alpha \right] \;=\; 
\pm\sigma\,\Tr(W^2 \Phi^2),\nonumber
\eea
where the upper/lower sign for $SO/Sp$ groups. Things are the same for the 
other pairs of (b), (e) and (c), (f) to yield the twisted half of $W^{(0,2)}$ 
as $\mp 2\sigma (1-\sigma)^2 \,\Tr(\wh{S}\Phi^2)$. Similarly one can 
evaluate the twisted half of $W^{(s,2n-s)}$. Since twisting reduces a double 
trace to a single trace, the twisted half has no dependence on $s$ to become 
$\mp 2\sigma (1-\sigma)^{2n}\,\Tr(\wh{S}\Phi^{2n})$. 
The holomorphic potential $\wt{W}$ splits into untwisted and twisted parts; 
$\wt{W}=\wt{W}_{un}+\wt{W}_{tw}$. Their $\Phi$-derivatives are given 
by 
\bea
\wt{W}^{\,\prime}_{un}
&=& 2\Phi -{1 \over 2}\oint_{\cal C} \frac{dz}{2\pi i}
\left[ {1 \over z-\Phi+i}+{1 \over z-\Phi+i} \right] 
\left[ R(z)+w^\alpha(z)\wh{v}_\alpha+T(z)\wh{S} \right],\nonumber\\
\wt{W}_{tw}^{\,\prime}
&=& \pm\sigma(1 -\sigma)\,
\wh{S}\left[ {1 \over 2\Phi+i}+{1 \over 2\Phi-i} \right].
\eea
Note that the twisted potential exists only for $\sigma=-1$.

The anomaly term in (\ref{gfloop}) shares the same one-loop structure 
as the holomorphic potential $\wt{W}$ so that the same modification is 
necessary to gain the correct $SO/Sp$ loop equations. 
Again, we replace $e_{ij}$ with $P_{ij}$ so that $P_{ij}$ acts on both 
$\delta \Phi_{kl} =f(\Phi,W_\alpha)_{kl}$ and $\partial/\partial\Phi_{ij}$. 
This modification was worked out in \cite{KRS}. The results are
\bea
\Tr \left[ \wt{W}^{\,\prime} \wh{R}(z) \right] 
&=& {1 \over 2}\,R(z)^2, \nonumber\\
\Tr \left[ \wt{W}^{\,\prime}\, \wh{T}(z) \right]
&=& \left \{ 
\begin{array}{ll}
\left( T(z) - \frac{2}{z} \right) R(z) & \quad SO^-, \\[8pt]
\left( T(z) - 2\frac{d}{dz} \right) R(z) & \quad SO^+, \\[8pt]
\left( T(z) + \frac{2}{z} \right) R(z) & \quad Sp^-, \\[8pt]
\left( T(z) + 2\frac{d}{dz} \right) R(z) & \quad Sp^+,
\end{array} \right.
\label{sosp1}
\eea
For $SO/Sp$ groups one can prove $w_\alpha (z) \equiv 0$ by using chiral 
ring properties \cite{KRS}. In ${\cal N}=1^*$ $SO/Sp$ theories, the 
holomorphic potential $\wt{W}$ also receives the corrections we figured out. 

Since $\wt{W}_{tw}^{\,\prime}$ is proportional to $\wh{S}$, this is irrelevant 
in the first equation in (\ref{sosp1}), i.e,
\beq
\Tr \left[ \wt{W}_{un}^{\,\prime} \wh{R}(z) \right] = {1 \over 2}\,R(z)^2.
\eeq
In the second equation, $\wt{W}_{tw}^{\,\prime}$ provides inhomogeneous 
corrections for $SO^- /Sp^-$ to yield  
\beq
\Tr \left[ \wt{W}_{un}^{\,\prime}\, \wh{T}(z) \right]
= \left \{ 
\begin{array}{ll}
\left[ T(z) - 2\left({1 \over z} -{1 \over 2z +i}-{1 \over 2z -i} \right)
\right]R(z) -\xi(z) & \quad SO^-, \\[8pt]
\left( T(z) - 2{d \over dz} \right)R(z) & \quad SO^+, \\[8pt]
\left[ T(z) + 2\left({1 \over z} -{1 \over 2z +i}-{1 \over 2z -i} \right)
\right]R(z) +\xi(z) & \quad Sp^-, \\[8pt]
\left( T(z) + 2{d \over dz} \right)R(z) & \quad Sp^+,
\end{array} \right.
\label{sosp2}
\eeq
where $\xi(z)$ is defined by
\beq
\xi(z) \equiv \frac{R(-{i \over 2})}{z+{i \over 2}}
+\frac{R(+{i \over 2})}{z-{i \over 2}}.
\eeq
The inhomogeneous corrections due to twisting reflect the informations of 
the matrix model resolvent defined on the ${\bf RP^2}$ Riemann surface.

\subsection{Connection to the matrix model loop equations}

The connection between the resolvents in field theories and those in matrix 
models was discussed in \cite {KRS}. Now in (\ref{floop2}) the first 
equation is closed by itself and coincides with the matrix model loop 
equation through the identification 
\beq
R(z) = {\bf R}_{S^2}(z), \label{RRS2}
\eeq
with the matrix model resolvent defined on a Riemann surface of $S^2$ 
(two-sphere). This identification can be understood as follows. 
Following \cite{DV1,DV2,KRS}, we define the matrix model resolvent 
${\bf R}_{S^2}(z)$ as
$$
{\bf R}_{S^2}(z) = {\bf g}\left\langle \Tr\,{1 \over z-{\bf \Phi}} 
\right\rangle
$$
with the matrix model coupling constant ${\bf g}$. 
The planar diagrams contributing to the glueball superpotential are the 
leading order terms in the $1/N$ expansion of matrix model free energy 
with the finite `t Hooft coupling  ${\bf S}={\bf g}N$. Then the prefactor 
${\bf g}$ in the matrix model resolvent is actually ${\bf S}/N$ so that 
the identification $S = {\bf S}$ of glueball superfield leads us to 
(\ref{RRS2}). 

According to \cite{KRS}, the other two resolvents are given by 
${\bf R}_{S^2}(z)$ as follows.
\bea
w_\alpha(z) &=& v_\alpha {\partial \over \partial {\bf S}}
{\bf R}_{S^2}(z), \nonumber\\
T(z) &=& N {\partial \over \partial {\bf S}}
{\bf R}_{S^2}(z) +{v^\alpha v_\alpha \over 2} 
{\partial^2 \over \partial {\bf S}^2}{\bf R}_{S^2}(z).
\label{resol}
\eea
This identification is obvious in (\ref{floop2}). 
One can see that the second and third equations in (\ref{floop2}) 
are satisfied by the identification above because the ${\bf S}$ dependence 
of the first equation is only through the resolvent $R(z)$. 
The identification in ${\cal N}=1^*$ theories was also examined 
in \cite{PTZ} and is consistent with \cite{KRS}. 

For $SO/Sp$ gauge groups, the planar diagrams of adjoint loops involve 
unoriented Riemann surfaces of ${\bf RP}^2$. The matrix model loop 
equations therefore include the resolvent ${\bf R}_{RP^2}(z)$ defined on 
the geometry of ${\bf RP}^2$. According to \cite{KRS}, the field theory 
resolvents are identified as follows.
\bea
R(z) &=& {\bf R}_{S^2}(z), \nonumber\\
T(z) &=& N {\partial \over \partial {\bf S}}
{\bf R}_{S^2}(z)+4\,{\bf R}_{RP^2}(z).
\eea
Obviously, the first term in $T(z)$ satisfies the homogeneous part of 
(\ref{sosp2}) so that the source of the inhomogeneity is 
nothing but the resolvent ${\bf R}_{RP^2}(z)$.  This lead us to predict that 
the matrix model loop equation for ${\bf R}_{RP^2}(z)$ might be written as
\bea
({\bf R}_{S^2} \circ {\bf R}_{RP^2})(z)
+({\bf R}_{RP^2} \circ {\bf R}_{S^2})(z)
&=& -2[V^{\,\prime}\,{\bf R}_{RP^2}(z)]_- \nonumber\\[8pt]&& 
\hspace{-60mm}+\left \{ 
\begin{array}{ll}
\left[ 2{\bf R}_{RP^2}(z) - \left({1 \over z} 
-{1 \over 2z +i}-{1 \over 2z -i} \right) \right] {\bf R}_{S^2}(z) 
-{1 \over 2}\xi(z) 
& \quad SO^-, \\[8pt]
\left( 2{\bf R}_{RP^2}(z) - {d \over dz} \right) {\bf R}_{S^2}(z) 
& \quad SO^+, \\[8pt]
\left[ 2{\bf R}_{RP^2}(z) + \left({1 \over z}
 -{1 \over 2z +i}-{1 \over 2z -i} \right) \right] {\bf R}_{S^2}(z) 
+{1 \over 2}\xi(z) 
& \quad Sp^-, \\[8pt]
\left( 2{\bf R}_{RP^2}(z) + {d \over dz} \right) {\bf R}_{S^2}(z) 
& \quad Sp^+, \\[8pt]
\end{array} \right.
\label{RP2}
\eea
by using the noncommutative product $(X \circ Y)$ peculiar to 
${\cal N}=1^*$ theories.

\setcounter{equation}{0}
\section{Conclusions}

In this paper, we examined the equivalence between the Konishi anomaly 
equations and the matrix model loop equations in ${\cal N}=1^*$ theories 
of classical gauge groups. For $U(N)$ gauge groups, we verified that 
the Konishi anomaly equations correctly reproduce the matrix model results 
in \cite{DV3,KKN}. Furthermore, we extended the field theory calculation to 
cover $SO/Sp$ cases also and predicted the matrix model loop equation 
(\ref{RP2}) for ${\bf R}_{RP^2}(z)$. This suggests that all the resolvents for 
possible Riemann geometries in the bosonic matrix model combine into a single 
resolvent equipped by an ${\cal N}=2$ glueball superfield $\cal S$ in 
field theory side. It might be interesting to analyze the loop equation 
(\ref{RP2}) by using the techniques of matrix models. 

For future direction, the field theoretical derivation of loop equations 
is based on the chiral rings among the chiral operators and is naturally 
extended to include the gravitational corrections from non-planar diagrams 
\cite{DGN,ACDGN,KMT}. It is well known that, generalizing the mass 
deformation into a polynomial potential $V(\Phi)$, the gauge theory dynamics 
of ${\cal N}=1^*$ involves the $SL(2,{\bf Z})$ action on $(p,q)$ confining 
vacua corresponding to $(p,q)$ 5-branes in string theory setup \cite{PS}. 
Now the gauge group is broken down to a product subgroup each 
semi-simple factor generates a branch cut on the complex $z$-plane. 
Such a multi-cut Riemann surface describes the Donagi-Witten spectral curve 
\cite{DW} and was studied in matrix model context in \cite{PTZ}, and in 
\cite{hollowood,hollowood2} describing the most general multi-cut solutions. 
It might be interesting 
to study such a gauge theory dynamics in ${\cal N}=1^*$ theory by analyzing 
the loop equations as performed in \cite{CSW,AO,KLLR,BINOR}.
Extension to the Leigh-Strassler deformation and comparing with the matrix 
model results in \cite{DHK,hollowood2} also seem interesting. 
Another interesting approach is to study ${\cal N}=1^*$ loop equations in 
relation to the large $N$ reduced model discussed in \cite{KKM}.

\vspace*{12pt}
\noindent
{\bf Acknowledgments:}
The author is grateful to Chanju Kim, Yoji Michishita and Hisao Suzuki 
for stimulating discussions. 
This work is the result of research activities (Astrophysical Research 
Center for the Structure and Evolution of the Cosmos (ARCSEC)) supported 
by Korea Science $\&$ Engineering Foundation. 

\appendix

\renewcommand{\thesection}{\large \bf \mbox{Appendix~}\Alph{section}}
\renewcommand{\theequation}{\Alph{section}\mbox{.}\arabic{equation}}

\section{\large \bf Computation of the one-loop effective action}
\setcounter{equation}{0}

The covariant derivatives we used in the text are 
\beq
D_\alpha =\partial_\alpha 
+i\sigma^m_{\alpha\dbe}\bar{\theta}^{\dbe}\partial_m,\quad
\oD^{\dalp} = \bar{\partial}^{\dalp} 
+i\bar{\sigma}^{m\dalp\beta}\theta_\beta \partial_m,\label{anticom}
\eeq
which are subject to the anti-commutation relation
$$
\{D_\alpha,\oD_{\dalp}\}=-2i\sigma^m_{\alpha\dalp}\partial_m.
$$
Useful identities are
$$
D^2 \oD^2 D^2 =-16 \,\square D^2,\quad
\oD^2 D^2 \oD^2 =-16 \,\square \oD^2.
$$

Now let us introduce the gauge covariant propagators
\beq
G_\pm ={-1 \over \square_\pm +m^2},\quad
K_\pm = {1 \over 16\,\square_\pm (\square_\pm +m^2)}.
\eeq
where 
\bea
\square_+ &\equiv& \square_{\rm cov}+{1\over 2}\cW^\alpha \nabla_\alpha
+{1\over 4}(\nabla^\alpha \cW_\alpha),\nonumber\\
\square_- &\equiv& \square_{\rm cov}-{1\over 2}
\overline{\cW}_{\dalp}\onab^{\dalp}
-{1\over 4}(\nabla^\alpha \cW_\alpha),\nonumber
\eea
with the gauge covariant d'Alembertian
$$
\square_{\rm cov} \equiv \square -2i\Gamma^m \partial_m 
-i(\partial_m \Gamma_m) -\Gamma^m \Gamma_m.
$$
Gauge connections are given by
$$
\Gamma^m \equiv -{1\over 4}\overline{\sigma}^{m \alpha\dalp}
\Gamma_{\alpha\dalp},\quad
\Gamma_{\alpha \dot{\alpha}} \equiv 
-i\{\nabla_\alpha,\onab_{\dalp}\}+i\{D_\alpha,\oD_{\dalp}\},\quad
\Gamma_\alpha \equiv \nabla_\alpha -D_\alpha.
$$
In the gauge chiral representation, the spinor connections become
$$
\Gamma_{\alpha \dot{\alpha}} \equiv \oD_{\dalp}\Gamma_\alpha,\quad
\Gamma_\alpha \equiv e^{-ad(V)}(D_\alpha e^{ad(V)}).
$$
Useful identities are
\beq
\nabla^2 \onab^2 \nabla^2 =-16 \,\square_- \nabla^2,\quad
\onab^2 \nabla^2 \onab^2 =-16 \,\square_+ \nabla^2.
\label{covid}
\eeq
\beq
\nabla^2 {1\over\square_+}\onab^2 ={1\over\square_-}\nabla^2
\onab^2,\quad
\onab^2 {1\over\square_-}\nabla^2 ={1\over\square_+}\onab^2\nabla^2.
\label{sliding}
\eeq
Using (\ref{sliding}), one can show that
\beq
\onab^2 \nabla^2 K_+ \onab^2 = G_+ \onab^2.\label{Ksliding}
\eeq

The holomorphic $2n$-point function is given by
\bea
\Gamma^{(2n)} &=& -{i(-)^n \over 2n}\,m^{2n}\,
\Tr_{\rm ad}\prod_{r=1}^{2n}\int_{z_r} 
\langle z_r|(\nabla^2 K_+ \onab^2)\phi|z_{r+1}\rangle \nonumber\\
&=& -{i(-)^n \over 2n}\,m^{2n}\,
\Tr_{\rm ad}\!\int_{z_1} \langle z_1|(\onab^2 \nabla^2 K_+)\phi|z_2\rangle
\prod_{r=2}^{2n}\int_{z_r} \langle z_r|G_+ \phi|z_{r+1}\rangle, \nonumber
\eea
where $\int_z \equiv \int d^8 z$ denotes the full superspace integral 
and in the second equation we have used the identities (\ref{covid}), 
(\ref{Ksliding}). $\sum_{n=1}^\infty \Gamma^{(2n)}$ gives 
the full one-loop correction to the effective action. 

Now we are going to expand the integrand with respect to the gauge 
connections to pick up the terms quadratic in $\cW_\alpha$. 
The gauge field dependence is only through the covariant vertex 
$\onab^2 \nabla^2$ and the gauge covariant propagators. 
Choosing the gauge chiral representation, $\onab_{\dalp}=\oD_{\dalp}$ 
and we are led to the expansion with respect to
$$
\delta\square_+ \equiv \square_+ -\square,\quad
\delta\nabla^2 \equiv \nabla^2 -D^2.
$$
By using the Schwinger-Dyson equations for the propagators, we see that
\bea
G_+ &=& G^{(0)}+G^{(1)}\delta\square_+ 
+G^{(2)}(\delta\square_+)^2 +\cdots,\nonumber\\
K_+ &=& K^{(0)}+K^{(1)}\delta\square_+ 
+K^{(2)}(\delta\square_+)^2 +\cdots, \nonumber
\eea
where
\bea
&& G^{(0)} = {-1\over \square +m^2},\quad 
G^{(1)} = (G^{(0)})^2,\quad G^{(2)} = (G^{(0)})^3,\nonumber\\
&&K^{(0)} = {1\over 16\,\square (\square +m^2)},\quad
K^{(1)} = K^{(0)}\left({-1\over \square}
+{-1\over \square +m^2}\right),\nonumber\\
&& K^{(2)} = K^{(0)}\left({1\over \square^2}
+{1\over \square(\square +m^2)}+{1\over (\square +m^2)^2}
\right). \nonumber
\eea
Then the relevant contributions are summarized as follows. 
\beq
\Gamma^{(2n)} = {(-)^n \over 2n}\left[
\Gamma^{(a)}+\Gamma^{(b)}+\Gamma^{(c)} +\sum_{r=2}^{2n}\left(
\Gamma^{(d)}_r+\Gamma^{(e)}_r+\Gamma^{(f)}_r \right)
+(1-\delta_{1n})\!\!\sum_{2\le r < s\le 2n}\!\!\Gamma^{(g)}_{rs}\right],
\label{2npoint}
\eeq
where the integrals are given by
\bea
\Gamma^{(a)} &=& -im^{2n}\!\int_{z_1}\!\dots \!\int_{z_{2n}}\!
\Tr_{\rm ad}\!\left[(\oD^2 \!D^2 K^{(1)}\delta\square_+)_1 
\hat{\delta}_{12} \phi_2 G_{(2,2n+1)}\right],\nonumber\\
\Gamma^{(b)} &=& -im^{2n}\!\int_{z_1}\!\dots \!\int_{z_{2n}}\!
\Tr_{\rm ad}\!\left[(\oD^2 \!D^2 K^{(2)}(\delta\square_+)^2)_1 
\hat{\delta}_{12} \phi_2 G_{(2,2n+1)}\right],\nonumber\\
\Gamma^{(c)} &=& -im^{2n}\!\int_{z_1}\!\dots \!\int_{z_{2n}}\!
\Tr_{\rm ad}\!\left[(\oD^2 \delta\nabla^2 K^{(1)}\delta\square_+)_1 
\hat{\delta}_{12} \phi_2 G_{(2,2n+1)}\right],\nonumber\\
\Gamma^{(d)}_r &=& -im^{2n}\!\int_{z_1}\!\dots \!\int_{z_{2n}}\!
\Tr_{\rm ad}\!\left[(\oD^2 \!D^2 K^{(1)}\delta\square_+)_1 
\hat{\delta}_{12} \phi_2 G_{(2,r)}
\right. \nonumber\\&& \left.\qquad\qquad\qquad\qquad\qquad
(G^{(1)}\delta\square_+)_r \hat{\delta}_{r,r+1}\phi_{r+1}
G_{(r+1,2n+1)}\right],\nonumber\\
\Gamma^{(e)}_r &=& -im^{2n}\!\int_{z_1}\!\dots \!\int_{z_{2n}}\!
\Tr_{\rm ad}\!\left[(\oD^2 \delta\nabla^2 K^{(1)}\delta\square_+)_1 
\hat{\delta}_{12} \phi_2 G_{(2,r)}
\right. \nonumber\\&& \left.\qquad\qquad\qquad\qquad\qquad
(G^{(1)}\delta\square_+)_r \hat{\delta}_{r,r+1}\phi_{r+1}
G_{(r+1,2n+1)}\right],\nonumber\\
\Gamma^{(f)}_r &=& -im^{2n}\!\int_{z_1}\!\dots \!\int_{z_{2n}}\!
\Tr_{\rm ad}\!\left[(\oD^2 \!D^2 K^{(0)}\delta\square_+)_1 
\hat{\delta}_{12} \phi_2 G_{(2,r)}
\right. \nonumber\\&& \left.\qquad\qquad\qquad\qquad\qquad
(G^{(2)}(\delta\square_+)^2)_r \hat{\delta}_{r,r+1}\phi_{r+1}
G_{(r+1,2n+1)}\right],\nonumber\\
\Gamma^{(g)}_{rs} &=& -im^{2n}\!\int_{z_1}\!\dots \!\int_{z_{2n}}\!
\Tr_{\rm ad}\!\left[(\oD^2 \!D^2 K^{(0)}\delta\square_+)_1 
\hat{\delta}_{12} \phi_2 G_{(2,r)}
\right. \nonumber\\&& \left.\qquad\qquad\qquad\qquad\qquad
(G^{(1)}\delta\square_+)_r \hat{\delta}_{r,r+1}\phi_{r+1} G_{(r+1,s)}
\right. \nonumber\\&& \left.\qquad\qquad\qquad\qquad\qquad\qquad
(G^{(1)}\delta\square_+)_s \hat{\delta}_{s,s+1}\phi_{s+1} G_{(s+1,2n+1)}
\right].\label{superF}
\eea
We have used the abbreviations;
$$
\hat{\delta}_{rs} \equiv \delta^8 (z_r-z_s), \quad
G_{(p,q)} \equiv \prod_{r=p}^{q-1}G^{(0)}_r 
\hat{\delta}_{r,r+1}\phi_{r+1}, \quad G_{(p,p)}\equiv 1.
$$
In the gauge chiral representation, one can see that
\bea
\delta\square_+ &=& {1 \over 2} \cW^\alpha D_\alpha 
+{1\over 4}(D^\alpha \cW_\alpha)
+{1\over 4}[\cW^\alpha,\Gamma_\alpha]
-2i\Gamma^m \partial_m -i(\partial^m \Gamma_m)-\Gamma^m 
\Gamma_m,\nonumber\\
\delta\nabla^2 &=& 2\Gamma^\alpha D_\alpha 
+(D_\alpha \Gamma_\alpha)+\Gamma^\alpha \Gamma_\alpha.\nonumber
\eea

In order to evaluate the integrals first we decompose the delta function 
such that
$$
\hat{\delta}_{rs}=\hat{\delta}^B_{rs}\hat{\delta}^F_{rs},
$$
where the bosonic and fermionic part are given by
$$
\hat{\delta}^B_{rs}\equiv\delta^4 (x_r-x_s),\quad
\hat{\delta}^F_{rs}\equiv\delta^4 (\theta_r-\theta_s)
\equiv(\theta_r-\theta_s)^2 (\bar{\theta}_r-\bar{\theta}_s)^2.
$$
Recall that, by using the anti-commutation relation (\ref{anticom}), 
one can reduce the number of active spinor derivatives to at most four.
The product of less than four D-operators in between two delta functions 
becomes zero, i.e, 
$$
\hat{\delta}^F_{21}\hat{\delta}^F_{12}=0,\quad
\hat{\delta}^F_{21}D_\alpha \hat{\delta}^F_{12}=0,\quad
\hat{\delta}^F_{21}D^2 \hat{\delta}^F_{12}=0,\quad
\hat{\delta}^F_{21}D_\alpha \oD^2 \hat{\delta}^F_{12}=0.
$$
and so on, so that for the Feynman integrals not to vanish, one has to 
pick up exactly two $D$'s and two $\oD$'s from $\delta\square_+$ and 
$\delta\nabla^2$. Then using the identity,
$$
\hat{\delta}^F_{21}\oD^2 \!D^2 \hat{\delta}^F_{12}=16\,
\hat{\delta}^F_{12},
$$
one is left with a single Berezin integral. 
In the last step, one must convert all the gauge connections into the gauge 
field strengths by using
\beq
\int\!d^4 x \,d^4\theta = -{1\over 4}\int\!d^4 x \,d^2\theta\,\oD^2,
\eeq
and the identities;
\beq
\cW_\alpha =-{1\over 4}(\oD^2 \Gamma_\alpha),\quad
\cW^\alpha \otimes \cW_\alpha =-{1\over 2}(\oD^2 \Gamma^m 
\otimes \Gamma_m).
\eeq

Finally, we perform the derivative expansion in each Feynman integral 
and take the lowest order term to reduce it to a local interaction term.
One can see that
\bea
\Gamma^{(a)} &=& 0,\nonumber\\
\Gamma^{(b)}+\Gamma^{(c)} &=& \wt{\Pi}_n \int\!d^4 x\,d^2\theta 
\,\Tr_{\rm ad}\!\left[ \cW^\alpha \cW_\alpha \phi^{2n} \right],\nonumber\\
\Gamma^{(d)}_r +\Gamma^{(e)}_r &=& \wt{\Pi}_n \int\!d^4 x\,d^2\theta 
\,\Tr_{\rm ad}\!\left[ \cW^\alpha \phi^{r-1} \cW_\alpha \phi^{2n-r+1}
\right],\nonumber\\
\Gamma^{(f)}_r &=& \wt{\Pi}_n \int\!d^4 x\,d^2\theta 
\,\Tr_{\rm ad}\!\left[ \cW^\alpha \cW_\alpha \phi^{2n} \right],\nonumber\\
\Gamma^{(g)}_{rs} &=& \wt{\Pi}_n \int\!d^4 x\,d^2\theta 
\,\Tr_{\rm ad}\!\left[ \cW^\alpha \phi^{s-r} \cW_\alpha \phi^{2n-s+r}
\right],
\eea
where the infrared divergences arising both in $\Gamma^{(b)}$ and 
$\Gamma^{(c)}$ cancel each other to give a finite result of the second 
equation. The coefficient $\wt{\Pi}_n$ is given by the Feynman integral 
\beq
\wt{\Pi}_n = {1\over 2}\int\!\!{d^4 q \over i(2\pi)^4}\,
m^{2n}\left({1\over q^2 -m^2}\right)^{2n+2} 
= {1\over 64\pi^2 m^{2n}}\left[{1\over n(2n+1)}\right].
\eeq
Substituting these results into (\ref{2npoint}), we obtain 
(\ref{holom}), (\ref{holom2}).


\end{document}